# Light Generation and Harvesting in a Van der Waals Heterostructure

Oriol Lopez-Sanchez[1], Esther Alarcon Llado[2], Volodymyr Koman[3], Anna Fontcuberta i Morral[2], Aleksandra Radenovic[4], Andras Kis[1*]

[1]*Electrical Engineering Institute, Ecole Polytechnique Federale de Lausanne (EPFL), CH-1015 Lausanne, Switzerland*

[2]*Institute of Materials, Ecole Polytechnique Federale de Lausanne (EPFL), CH-1015 Lausanne, Switzerland*

[3]*Institute of Microtechnology, Ecole Polytechnique Federale de Lausanne (EPFL), CH-1015 Lausanne, Switzerland*

[4]*Institute of Bioengineering, Ecole Polytechnique Federale de Lausanne (EPFL), CH-1015 Lausanne, Switzerland*

*\*Correspondence should be addressed to: Andras Kis, andras.kis@epfl.ch*

## ABSTRACT

Two-dimensional (2D) materials are a new type of materials under intense study because of their interesting physical properties and wide range of potential applications from nanoelectronics to sensing and photonics. Monolayers of semiconducting transition metal dichalcogenides $MoS_2$ or $WSe_2$ have been proposed as promising channel materials for field-effect transistors (FETs). Their high mechanical flexibility, stability and quality coupled with potentially inexpensive production methods offer potential advantages compared to organic and crystalline bulk semiconductors. Due to quantum mechanical confinement, the band gap in monolayer $MoS_2$ is direct in nature, leading to a strong interaction with light that can be exploited for building phototransistors and ultrasensitive photodetectors. Here, we report on the realization of light-emitting diodes based on vertical heterojunctions composed of n-type monolayer $MoS_2$ and p-type silicon. Careful interface engineering allows us to realize diodes showing rectification and light emission from the entire surface of the heterojunction. Electroluminescence spectra show clear signs of direct excitons related to the optical transitions between the conduction and valence bands. Our pn diodes can also operate as solar cells, with typical external quantum efficiency exceeding 4%. Our work opens up the way to more sophisticated optoelectronic devices such as lasers and heterostructure solar cells based on hybrids of two-dimensional (2D) semiconductors and silicon.

Molybdenum disulphide (MoS$_2$) is a typical representative of layered transition-metal dichalcogenide (TMD) semiconductors[1] with electronic properties and potential range of applications complementary to those of graphene. Bulk TMD crystals are stacks of layers held together via weak van der Waals interaction, allowing the extraction of single 2D atomic layers using either the adhesive-type based micromechanical cleavage technique[2] originally developed for the preparation of graphene. Because it has a band gap, monolayer MoS$_2$ can be used as the basic building block of room-temperature field-effect transistors[3] with an on/off ratio exceeding $10^8$ as well as logic circuits[3] and amplifiers[4] with high gain. Large-area MoS$_2$ can also be grown using CVD-like growth techniques[5,6] or deposited using liquid phase exfoliation.[7-9]

The electronic and optical properties of monolayer MoS$_2$ and other semiconducting dichalcogenides are fundamentally different from those of their bulk counterparts. Because of the lack of inversion symmetry, charge carriers in monolayer MoS$_2$ behave as massive Dirac fermions[10] while the conduction band of MoS$_2$ shows strong spin-orbit-induced spin splitting[11] and strong coupling of spin and valley degrees of freedom that can be detected using circularly polarized light[12-14] and could be used in novel devices based on valley Hall effect.[15]

A transition from an indirect band gap to a direct band gap occurs in the monolayer limit,[16-19] manifesting itself in strong photoluminescence.[17-18] The direct band gap in $MoS_2$ can also be harnessed for the realization of vertical optoelectronic devices [20] as well as phototransistors[21] and photodetectors[22] with high responsivity and low noise-equivalent power.[22] Sundaram et al. recently demonstrated[23] that monolayer $MoS_2$ can also be used as a light emitter in an electroluminescent device with light emission occurring due to hot carrier processes in a region near locally gated contacts. While this result showed that monolayer $MoS_2$ could be used for the fabrication of light-emitting devices, the device geometry was limited by a relatively high power threshold for light emission and only a small portion of the device, restricted to the contact edge was active in electroluminescence.

One way to overcome these factors limiting the exploitation of monolayer $MoS_2$ for practical applications in optoelectronic devices is to build light-emitting diodes based on vertical pn junctions, resulting in a natural increase of the junction area which can easily be scaled. We demonstrate here such a junction in the form of a vertical heterostructure composed of n-type $MoS_2$ and p-type silicon Si serving as the hole injection layer.[24] We choose p-type silicon for this purpose because it is readily available and easy to pattern and handle. No reports on p-type monolayer $MoS_2$ have been published so far. Our device shows a decreased threshold power for light emission, while the entire heterojunction surface is active as a light emitter. The device is also capable of operating as a solar cell.

**Results and Discussion**

Figure 1 shows the structure of our device. Fabrication starts with exfoliation of $MoS_2$ (ref 2) onto an $SiO_2$/Si substrate. $MoS_2$ is then transferred[25] onto a pre-patterned target highly doped p-type Si substrate[26] covered with $SiO_2$ with 1 μm x 1 μm to 100 μm ×100 μm windows through which the underlying Si is exposed. The native oxide on the substrate is removed and the Si surface passivated with hydrogen using a second wet etch step.[27]

In order to avoid degradation of the passivation layer, monolayer $MoS_2$ is then immediately transferred across the edge of a window exposing the Si surface (figure 1a) and contacted on one side with a gold electrode. Based on AFM imaging we can see that $MoS_2$ is transferred on top of H-Si in a conformal fashion, with no visible voids or wrinkles. Both the 2D $MoS_2$ film and the H-Si substrate are terminated and have no dangling bonds at their surfaces, allowing the formation of a van der Waals heterostructure.[28] Because the nature of the interface is similar to that in graphene/BN heterostructures, we expect that most if the interface in our device is clean and free of contaminants[29] allowing direct charge injection between Si and $MoS_2$. For diode characterization and electroluminescence measurements, we use the measurement scheme shown on figure 1b. On some of the devices we also deposit a 30 nm thick $HfO_2$ or $Al_2O_3$ layer in order to encapsulate the device and increase the mobility in monolayer $MoS_2$.[30] This includes both devices presented in this manuscript. More than 10 functioning devices were produced showing similar characteristics.

Figure 2a shows the current vs. bias voltage (*I-V* ) characteristic of our $MoS_2$/Si heterojunction diode with 30 nm $HfO_2$ on top, exhibiting rectifying behavior with a current of 346 nA for a forward bias of 10 V and a junction area of 19 μm$^2$. This shows that classical diodes and all related optoelectronic devices could be prepared using a combination of an atomically thin 2D semiconductor and a 3D semiconductor which should allow for a rapid fabrication and development of this type of devices on industrial scales. We find that in the reverse bias regime a breakdown does not occur below -10V.

On figure 2b we outline the proposed band structure of our device, typical of type-II abrupt heterojunctions.[31] The junction is characterized by conduction ($\Delta E_C$ = 200 meV) and valence ($\Delta E_V$ = 900 meV) band offsets due to different electron affinities[32] and band gaps of Si and $MoS_2$. Under the application of forward bias $V$ to the heterojunction, electrons injected from the $MoS_2$ side and holes injected from p-Si can radiatively recombine in the junction, resulting in light emission. Due to the direct band gap nature of $MoS_2$,[16-19] we expect the emitted light to be characterized by radiative transitions in $MoS_2$ as radiative transitions in Si are expected to be much less efficient due to its indirect band gap. Due to valence and conduction band offsets, discontinuities could occur in the bands with valence and conduction band cusps that can impair charge carrier injection efficiency and limit the device current.

The electroluminescent emission intensity map for a forward bias of 10 V is shown on figure 2c, superposed on the outline of the device. Most of the heterojunction surface is active, in contrast to $MoS_2$ electroluminescent devices based on hot carrier processes in a region near locally gated contacts[23] or previously reported observations in a similar device geometry,[33] where the light emission was localized only at the heterojunction edge. The presence of a large active area in our device can be attributed to hydrogen passivation of the Si substrate, resulting in the formation of a true heterojunction with an efficient charge transfer. Large-area emitters such as the one presented here are also more attractive from the practical point of view because the total emitted light intensity could be more easily scaled up by simply increasing the device area.

Figure 2d shows the integrated electroluminescence intensity as a function of device current and electrical power for the active area centered on the heterojunction surface. The results show light emission from the device for bias voltages exceeding 5.5 V, corresponding to an electroluminescence threshold current of ~109 nA. The equivalent threshold power density is 3.2 W/cm$^2$, significantly lower than the previously reported threshold power of 15 kW/cm$^2$ for MoS$_2$ electroluminescent devices based on hot carrier injection.[23] This shows that the defect-free vertical heterojunction geometry and Si/MoS$_2$ band alignement are favorable for reducing the emission threshold and increasing the surface area of the emitter. The presence of a threshold is probably due to the existence of cusps in the heterojunction band diagram under forward bias conditions (figure 2b) and could probably be further decreased with careful band engineering of the interface.

The photoluminescence (PL) spectrum of monolayer MoS$_2$ is acquired in the region where the flake is supported by SiO$_2$ and is shown on figure 3a. The spectrum shows two peaks at 685 nm (1.81 eV) and 624 nm (1.98 eV). They are associated with excitonic transitions between the bottom of the conduction band and top of the valence band, split due to spin-orbit coupling.[17,18,11]

On figure 3b we show the electroluminescence spectrum, together with a fit to a multiple peak Lorentzian model. The main feature of the spectrum is a peak with a position of 694 nm (1.78 eV) which has a full width at half maximum of 56 nm. The position of this peak matches well with the observed PL peak at 685 nm and is associated with the A exciton[12, 17-18] in monolayer MoS$_2$. This shows that the relevant energy for the radiative recombination process in the MoS$_2$/Si heterojunction is the direct band gap in monolayer MoS$_2$. We observe an additional feature at 721 nm (1.72 eV) which can be related to the trion (negatively charged exciton) resonance in monolayer MoS$_2$.[34]

In addition to the A exciton, because of the low emission threshold, the electrical power density at which our device operates provides enough energy through impact ionization to excite the higher energy B exciton which we can distinguish as an additional feature in the electroluminescence spectrum, located at 644 nm (1.92 eV). The small red shift with respect to the related PL peak at 624 nm could be attributed to differences in the dielectric environments: the EL peak is acquired in the heterojunction region where MoS$_2$ is in direct contact with silicon while the PL peak is acquired in the region where MoS$_2$ is supported by SiO$_2$. The difference in these two dielectric environments could affect the exciton binding energy through screening of the Coulomb interaction between electrons and holes.

The photoluminescence is strongly quenched in the heterojunction area, indicating the presence of an internal, open circuit voltage that separates the electrons and holes. This is favorable for operating our van der Waals heterostructures as solar cells. On figure 4a we show current as a function of voltage in a second device with an area of 8 μm$^2$ for different illumination powers, showing the increase of short-circuit current under illumination, indicating power generation. The electrical power $P$ generated in the device defined as $P = I \times V$ is shown on figure 4b, with a peak power of ~1nW for a bias voltage V = ~0.1 V and illumination power of 861 nW. We characterize the spectral response of the solar cell by measuring its short circuit current $I_{sc}$ (obtained for a bias voltage of 0 V) as a function of illumination wavelength λ using a supercontinuum light source. The external quantum efficiency

*EQE* of the device, defined as $EQE = (I_{sc}/P_{inc}) \times (hc/e\lambda)$ where *h* is the Planck's constant, *c* the speed of light and *e* the elementary charge is shown on figure 4c. It is characterized by a sharp drop above 1000 nm, coinciding with the absorption edge of silicon and a broadband response in the 500-1000 nm region, indicating that $MoS_2$ and Si form a true pn heterojunction instead of a Schottky contact and operate in tandem. The complementary absorption profiles of these materials result in a device with spectral response that is extended with respect to the response of monolayer $MoS_2$.[35] The maximum recorded *EQE* is 4.4% and is a promising number for a device based on a two-dimensional monolayer and is more than an order of magnitude higher than in lateral pn junctions based on the dichalcogenide $WSe_2$ which also shows a much narrower spectral response limited by its band gap.[36,37] The *EQE* could be further enhanced by careful control over doping levels of the $MoS_2$ and Si that would reduce the series resistance of the device, use of large-area grown or deposited materials[7,6, 38] and by incorporating additional 2D semiconducting layers such as $WSe_2$ with complementary absorption spectra.

**Conclusion**

To summarize, we show electroluminescent devices and solar cells based on heterojunctions composed of monolayer $MoS_2$ and p-type silicon. This choice of materials combines the advantages of the direct band gap and small thickness of 2D $MoS_2$ with the established silicon-based fabrication processes and could show the way to implementing 2D semiconductors as enabling materials in standard semiconductor fabrication lines. Furthermore, all the semiconducting materials used in our devices can be considered earth abundant and non-toxic. The entire junction area in our device participates in light emission with a low emission threshold power, allowing future large-area light emitters and lasers based on $MoS_2$. The low threshold power allows us to distinguish features in the emitted light spectra related to three different optical transitions – A and B excitons and the A- trion resonance[34] which could find valuable applications in the field of valleytronics. The heterojunction diode can also operate as a photovoltaic device, converting incoming light into electrical power with an external quantum efficiency of 4.4% and a broad spectral response indicating that $MoS_2$ and silicon operate in tandem.

**METHODS**

Single layers of $MoS_2$ are exfoliated from commercially available crystals of molybdenite (SPI Supplies Brand Moly Disulfide) using the scotch-tape micromechanical cleavage technique method pioneered for the production of graphene[39] onto silicon substrates covered by 270 nm layer of thermal oxide. Monolayer samples were identified by optical microscopy.[40] Once identified, monolayers were transferred[25] onto p-type silicon substrates with a resistivity of 0.1-0.5 Ωcm, corresponding to a boron doping level between $3\times10^{16}$ and $3\times10^{17}$ $cm^{-3}$, covered by 100 nm thick layer of thermal $SiO_2$ with patterned holes from 1 μm x 1 μm up to 100 μm × 100 μm. Windows in  in $SiO_2$ are opened using 7:1 buffered oxide etch, resulting in sloped sidewalls. The initial etching step was followed 1 minute 1% HF etch in order to remove the native oxide and passivate the Si surface. [27] The sample thickness was confirmed by photoluminescence measurements.

Monolayer $MoS_2$ diodes were characterized at room temperature. For electrical characterization, we use a gold electrode deposited on MoS2 and a large-area electrodes in direct contact with the p-Si substrate. A second gold electrode is deposited on top of Si near the $MoS_2$ flake but not in direct electrical contact with it. We use this electrode to verify that charge carriers can be injected from the passivated Si substrate. The emitted radiation was collected and analyzed using a grating spectrometer (HORIBA Jobin Yvon) equipped with a liquid nitrogen cooled CCD camera (Triax 550). An Andor iXon Ultra camera was used to perform photon counting and map the light emission. Photoluminescence measurements were performed using a laser centered at 488 nm and a spectrometer (Princeton Instruments SP-2500i) with a liquid nitrogen cooled camera (PiXIS/Pylon/Spec-10:256).

Heterojunction band structures were based on modeling performed using the Adept 2.0 tool available on NanoHUB.org.

## ACKNOWLEDGEMENTS

Device fabrication was carried out in part in the EPFL Center for Micro/Nanotechnology (CMI). Thanks go to Zdenek Benes for technical support electron-beam lithography. We thank Olivier Martin (EPFL) for the use of the setup for the spectral characterization of electroluminescence as well as Jacopo Brivio and Simone Bertolazzi for technical help with the $MoS_2$ transfer. This work was financially supported by the Swiss Nanoscience Institute (NCCR Nanoscience).

**FIGURES**

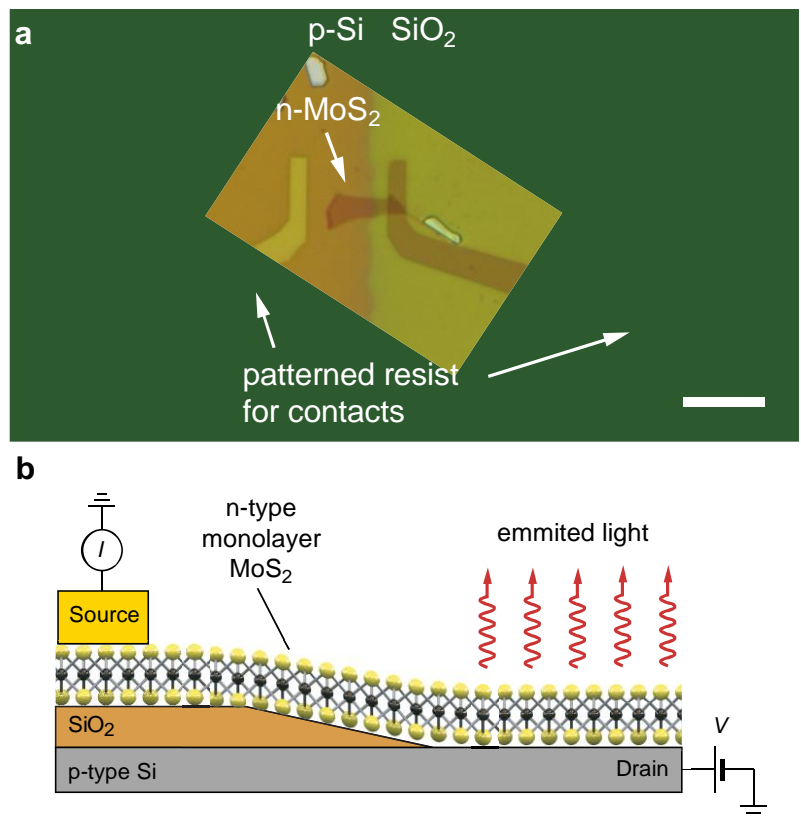

**Figure 1. The geometry of the MoS$_2$/Si heterojunction light-emitting diode. a,** Optical image of the device in an intermediate state of fabrication. Monolayer MoS$_2$ is placed across the sidewall of a square window etched into a SiO$_2$ layer and exposing the underlying p-doped silicon. Scale bar is 10 $\mu$m long. **b,** Cross-sectional view of the structure of the device structure together with electrical connections used to induce light emission from the heterojunction. Electrons are injected from n-type MoS$_2$ while holes are injected from the p-Si substrate.

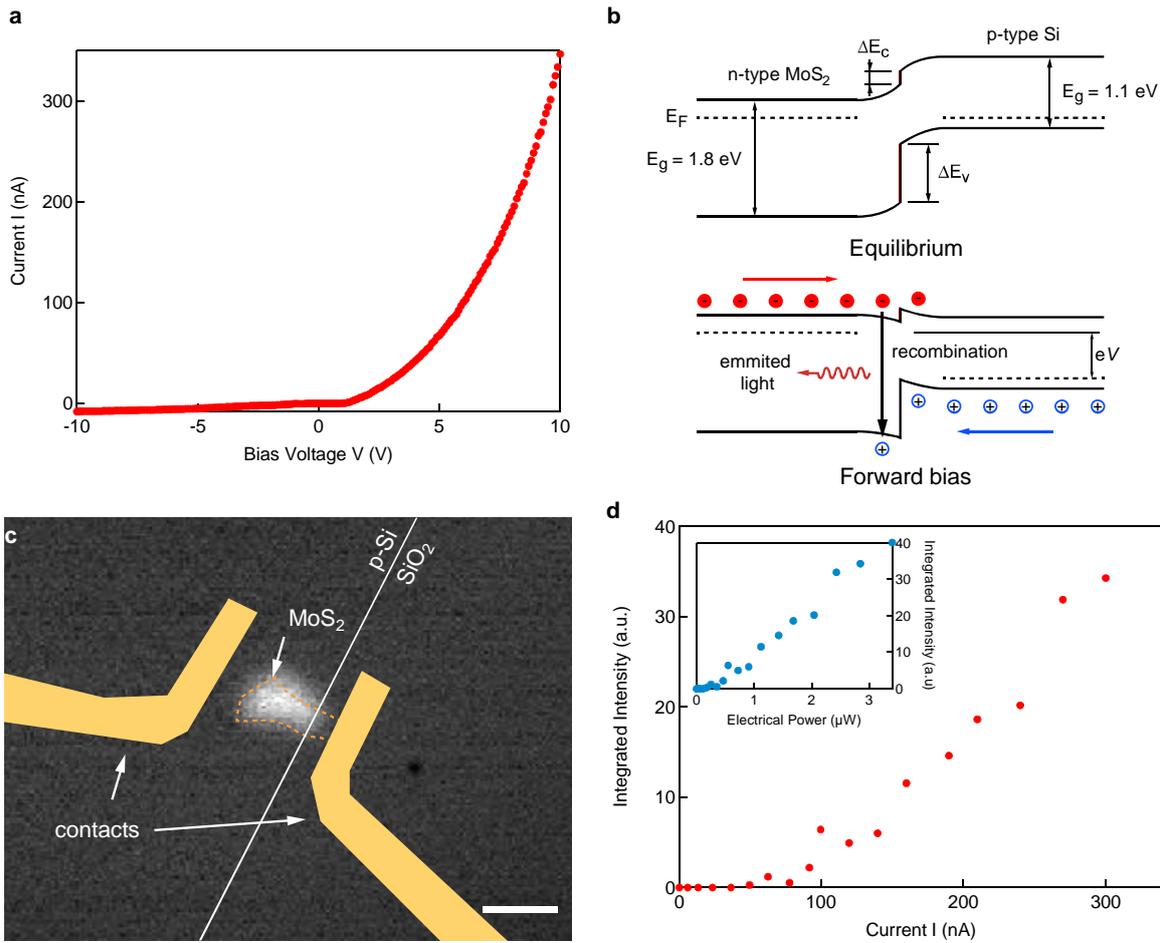

**Figure 2. Electrical characteristic of the device. a**, Current vs. bias voltage characteristic of the MoS₂/Si heterojunction diode. **b,** Band diagram of the MoS₂/Si heterojunction in equilibrium conditions and under forward bias. Electrons injected from the n-MoS₂ and holes from p-Si can radiatively recombine in the junction. **c,** Intensity map of showing the electroluminescent emission with superimposed outline of the most important device components. The entire surface of the heterojunction is emitting light. Scale bar is 5 μm long. **d,** Integrated light intensity as a function of device current. The inset shows the emitted light intensity as a function of electrical power. The threshold current for light emission is ~100 nA, corresponding to a threshold power of 3.2 W/cm² for a device with an active area of 19 μm².

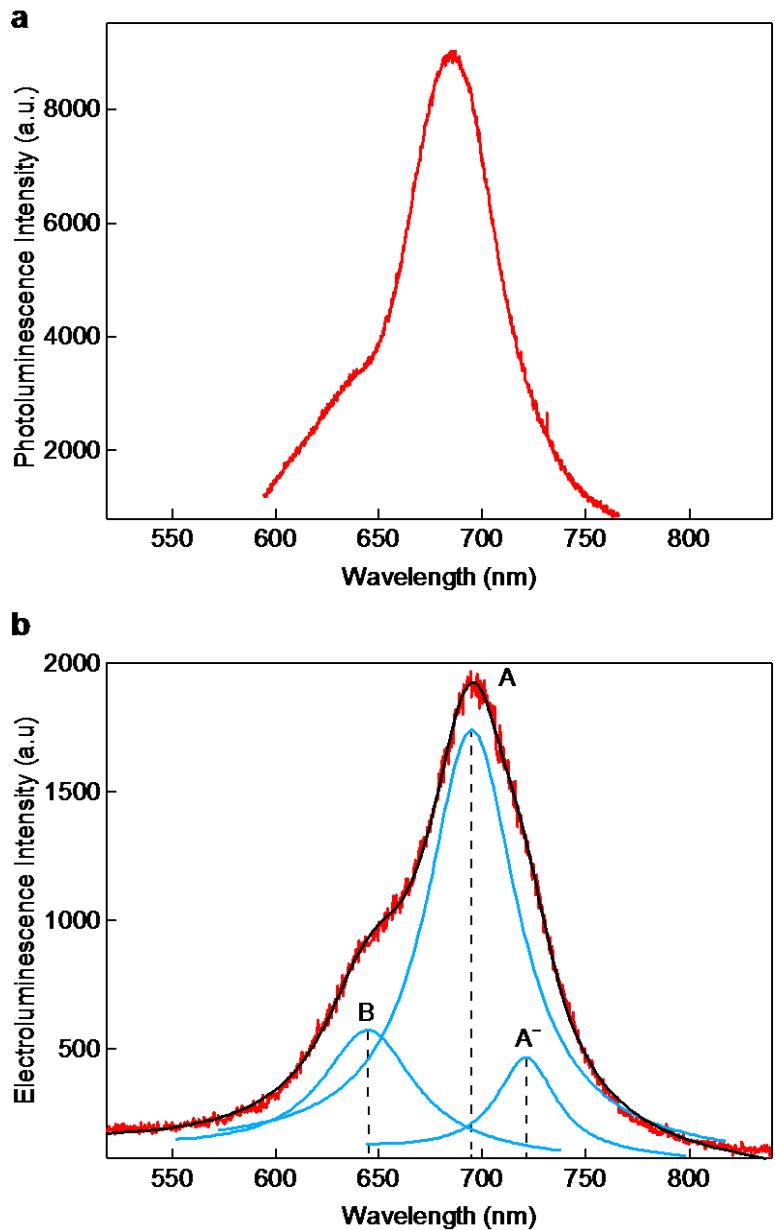

**Figure 3. Light emission characteristics of the device. a,** Photoluminescence spectrum of the region of monolayer $MoS_2$ flake supported by $SiO_2$. **b,** Electroluminescence spectrum acquired under a forward bias V = 15 *V* and a current of 1.8 μA. The spectrum is fitted with three Lorentzian lines which correspond to A and B excitons at 694 nm and 644 nm and the A$^-$ trion resonance at 721 nm.

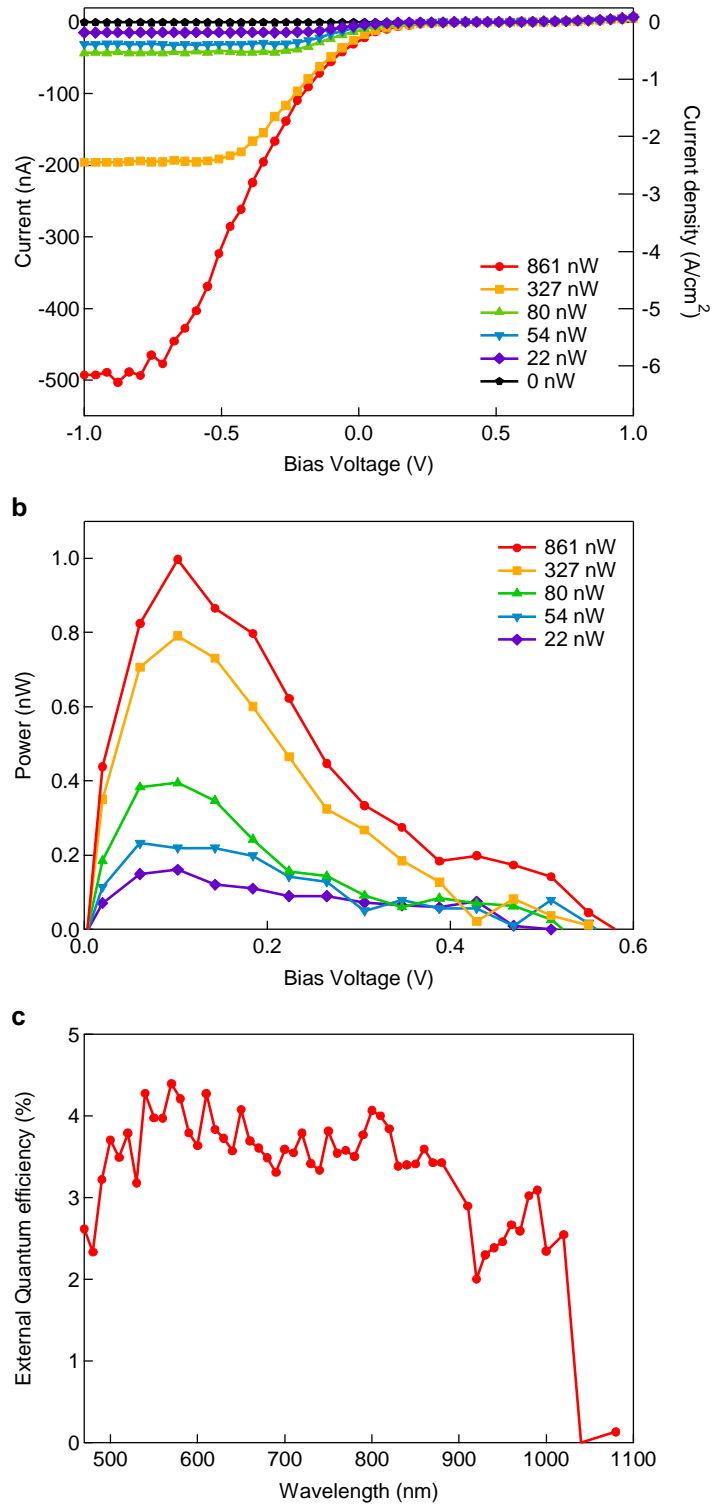

**Figure 4. MoS₂/Si heterojunction as a solar cell. a,** Current as a function of bias voltage under different illumination powers from a 541 nm laser. The heterojunction area is 8 μm². **b**, Electrical power generated by the device as a function of bias voltage, recorded for different illumination powers, extracted from data shown in a. **c**, External quantum efficiency as a function of wavelength in the 500-1100 nm range for an illumination power of  500 nW. The curve shows a broadband response with MoS₂ and Si working in tandem and effectively extending the spectral response of MoS₂ into the infrared region. At both ends of the wavelength range, our measurements are limited by the sharp drop in emission intensity of our supercontinuum light source.